\documentclass[9pt, conference]{IEEEtran}
\IEEEoverridecommandlockouts
\usepackage{cite}
\usepackage{url}
\usepackage{amsmath,amssymb,amsfonts}
\usepackage{algorithmic}
\usepackage{multirow}
\usepackage{booktabs}
\usepackage{arydshln}
\usepackage{array}
\usepackage[dvipdfmx]{graphicx}
\usepackage{textcomp}
\usepackage{xcolor}

\def\BibTeX{{\rm B\kern-.05em{\sc i\kern-.025em b}\kern-.08em
    T\kern-.1667em\lower.7ex\hbox{E}\kern-.125emX}}
\begin{document}

\title{Benchmarking Prosody Encoding in Discrete Speech Tokens
}

% \author{Anonymous submission to ASRU 2025}
\author{\IEEEauthorblockN{Kentaro Onda\IEEEauthorrefmark{1}\IEEEauthorrefmark{2}, Satoru Fukayama\IEEEauthorrefmark{2}, Daisuke Saito\IEEEauthorrefmark{1}, Nobuaki Minematsu\IEEEauthorrefmark{1}}
\IEEEauthorblockA{
\IEEEauthorrefmark{1}\textit{The University of Tokyo, Japan}
\IEEEauthorrefmark{2}\textit{National Institute of Advanced Industrial Science and Technology (AIST), Japan}\\
\{ondakentaro, dsk\_saito, mine\}@gavo.t.u-tokyo.ac.jp, s.fukayama@aist.go.jp
}
}

% \author{\IEEEauthorblockN{Kentaro Onda}
% \IEEEauthorblockA{\textit{Graduate School of Engineering} \\
% \textit{The University of Tokyo}\\
% Tokyo, Japan \\
% ondakentaro@gavo.t.u-tokyo.ac.jp}
% \and
% \IEEEauthorblockN{Satoru Fukayama}
% \IEEEauthorblockA{\textit{Artificial Intelligence Research Center} \\
% \textit{National Institute of Advanced Industrial}\\
% \textit{Science and Technology (AIST)}\\
% Tokyo, Japan \\
% s.fukayama@aist.go.jp}
% \and
% \IEEEauthorblockN{Daisuke Saito}
% \IEEEauthorblockA{\textit{Graduate School of Engineering} \\
% \textit{The University of Tokyo}\\
% Tokyo, Japan \\
% dsk\_saito@gavo.t.u-tokyo.ac.jp}
% \and
% \IEEEauthorblockN{Nobuaki Minematsu}
% \IEEEauthorblockA{\textit{Graduate School of Engineering} \\
% \textit{The University of Tokyo}\\
% Tokyo, Japan \\
% mine@gavo.t.u-tokyo.ac.jp}
% }

\maketitle

\begin{abstract}
Recently, discrete tokens derived from self-supervised learning (SSL) models via k-means clustering have been actively studied as pseudo-text in speech language models and as efficient intermediate representations for various tasks. However, these discrete tokens are typically learned in advance, separately from the training of language models or downstream tasks. As a result, choices related to discretization, such as the SSL model used or the number of clusters, must be made heuristically. In particular, speech language models are expected to understand and generate responses that reflect not only the semantic content but also prosodic features. Yet, there has been limited research on the ability of discrete tokens to capture prosodic information. To address this gap, this study conducts a comprehensive analysis focusing on prosodic encoding based on their sensitivity to the artificially modified prosody, aiming to provide practical guidelines for designing discrete tokens.
\end{abstract}


\section{Introduction}
Self-supervised learning (SSL) models \cite{hubert, wav2vec, Chen2021WavLMLS, Chung2021w2vBERTCC, sslreview} 
have demonstrated strong performance across various speech processing tasks 
and have been widely adopted in the field. Recently, there has been growing interest in 
discretizing their outputs for more effective utilization \cite{guo2025recentadvancesdiscretespeech}.
By discretizing speech, these tokens can be seen as ``pseudo-text" and
now (large) language models can process speech directly without
transcribing it into text \cite{lakhotiaetal2021generative,zhang2023speechgpt, borsos2023audiolm, rubenstein2023audiopalm,arora2025landscapespokenlanguagemodels}.
In addition to this, since each frame is represented by a single integer,
these tokens have attracted attention as a highly efficient alternative to
conventional continuous features in terms of data size and computational cost \cite{chang24b_interspeech}.
Their use as intermediate representations is also being explored in various tasks,
such as automatic speech recognition (ASR),
text-to-speech (TTS), and speech translation (ST) \cite{chang23b_interspeech, chang2024exploring,yang2024towards,mousavi24_interspeech,shi24h_interspeech}.

These discrete tokens are typically obtained by applying k-means clustering to 
SSL features. However, since downstream tasks are trained separately after 
the tokenization process, the resulting tokens are not necessarily optimal for the target tasks. 
As a result, the choice of the upstream SSL model, 
the layer to extract features from, and the number of clusters (`k' in k-means) 
must be determined heuristically.
It is also known that downstream task performance can be affected 
by post-processing for token sequences such as deduplication (removing repeated tokens)
and subword modeling (byte pair encoding) \cite{chang23b_interspeech,chang2024exploring,shen2024}. 
In addition, preprocessing applied to SSL features, such as independent component analysis (ICA) \cite{nakamura2025discretespeechunitextraction}
can also improve the downstream performance.
However, evaluating token quality based on downstream task performance is 
computationally inefficient, when aiming to explore better token designs.
Moreover, assesing token characteristics via downstream task performance entangles
the effects of the tokens themselves with those of the decoder in the downstream model,
making interpretation challenging.
Therefore, it is desirable to evaluate the properties of the tokens 
at the stage when only tokenization has been learned before training the downstream model.

Existing evaluations of tokens have largely focused on linguistic information, 
with many studies measuring robustness against acoustic varieties,
such as speaker identity, noise, and reverberation \cite{gat-etal-2023-augmentation,vashishth2024stabspeechtokenizerassessment,yeh2024estimating}. 
In these studies, pitch shifting and time stretching were also 
treated as acoustically redundant information that should be removed from 
discrete tokens, and evaluation metrics were designed accordingly. 
However, whether these prosodic information such as pitch, duration, and intensity 
should be included in or excluded from the tokens is still an open question.
% There is a notable lack of research that distinguishes between 
% paralinguistic and non-linguistic information 
% and evaluates their encoding.
In real human dialogue, paralinguistic information plays a crucial role.
For example, we often emphasize certain words or phrases to convey our intentions more clearly
with a higher pitch and a louder volume.
We may also change our speaking style to express emotions.
Speaking in a flat tone may convey a neutral or serious attitude,
while using dynamic intonation can indicate excitement or enthusiasm.
This kind of variations in prosody impart additional implicit meaning to the utterances
even if the linguistic content remains unchanged.
In tasks that require capturing the more comprehensive meaning of 
spoken language, discrete tokens need to be sensitive not only 
to linguistic information but also to paralinguistic information.
Given this context, it is essential to evaluate the properties of discrete tokens
with a clear distinction between paralinguistic and non-linguistic acoustic information.
Therefore, this study conducts a benchmarking analysis 
with a particular focus on prosody, 
aiming to evaluate and analyze how 
linguistic, paralinguistic, and non-linguistic information 
are encoded in discrete tokens.
The analyses were primarily conducted based on the sensitivity of the discrete tokens to artificially induced variations in prosody.
Our findings serve as a useful guideline for designing discrete tokens
tailored to specific objectives.

Our main findings are summarized as follows:
\begin{itemize}
    \item \textbf{Sensitivity to relative prosodic changes:} SSL models trained with frame-wise masked prediction responded more sensitively when only specific words were emphasized, compared to when the entire utterance was uniformly emphasized. This suggests that these models capture relative prosodic contours. This tendency was particularly prominent in the case of intensity.
    \item \textbf{Effect of masked prediction targets in SSL pretraining:} Models that use discrete labels as targets for masked prediction during pretraining tend to produce tokens that capture linguistic information even with a small number of clusters. In contrast, models that predict continuous features do not perform as well with small cluster sizes, and begin to capture linguistic structure more effectively only when the number of clusters is large, i.e., when the tokens approximate continuous representations. However, these model-specific tendencies are primarily observed near the final layers, while the intermediate layers exhibit broadly similar patterns across different models.
    \item \textbf{Training data for k-means clustering:} It was found that training k-means clustering with emotional speech improves the sensitivity to prosodic features of the obtained tokens, even with the same upstream SSL models.
    \item \textbf{Effect of applying a moving average to SSL features:} It was found that by applying a moving average with an appropriate window size to SSL features as a preprocessing, it is possible to improve both speaker invariance and sensitivity to prosody.
\end{itemize}

\section{Related Work}
\subsection{Acoustic Tokens vs. Semantic Tokens}
Discrete tokens are generally categorized into two types:
acoustic tokens and semantic (or phonetic) tokens \cite{guo2025recentadvancesdiscretespeech,arora2025landscapespokenlanguagemodels}.
Acoustic tokens are derived from neural audio codec models,
which aim to compress and accurately reconstruct the original audio \cite{zeghidour2021soundstream,fossez2023high}.
As such, they are designed to capture all the fine-grained details of the speech signal, 
including speaker identity and background noise, which are not related to what the speaker intends to convey.
In contrast, semantic tokens are obtained by clustering SSL features,
with the goal of capturing the linguistic content of speech \cite{lakhotiaetal2021generative}.
Indeed, semantic tokens tend to perform better in tasks where linguistic information is critical,
such as ASR \cite{yang2024towards,shi24h_interspeech}. In speech language models as well, semantic tokens are often used as 
a core substitute for textual information, whereas acoustic tokens are typically employed 
to embody the generated audio more richly \cite{borsos2023audiolm, rubenstein2023audiopalm}.
Therefore, in order to build a speech language model that can understand the implicit meanings
conveyed through prosody, it is essential for semantic tokens to be sensitive to prosodic information. 
While several studies have extensively benchmarked acoustic tokens \cite{shi2024espnet,wu-etal-2024-codec},
this study primarily focuses on semantic tokens for in-depth analysis of the prosody encoding.

\subsection{Prosodic information in Semantic Tokens}
As discussed in \cite{yeh2024estimating}, it has often been assumed without sufficient verification
that the discretization process of SSL features can remove speaker information from tokens
while retaining only linguistic information.
However, \cite{yeh2024estimating} demonstrated that speaker information remains sufficiently encoded in discrete tokens and is accessible.
\textit{What about prosodic information?} 
Prosody is considered to lie somewhere between linguistic information and speaker (acoustic) information,
as a partially embodied form of textual content.
Yet, while some studies claim that prosodic features are retained in semantic tokens \cite{guo2024vec2wav}, 
others argue they are not \cite{osakuade2024discrete, kharitonov-etal-2022-text}.
We believe the reason of this controversy is that such analyses mainly rely on the performance of downstream tasks,
such as voice conversion and tone classification.
Since the performance depends heavily on the training and evaluation data, as well as 
the decoder's capabilities, it does not provide a direct understanding of the prosody encoding in discrete tokens.
Therefore, in this study, we aim to analyze the properties of discrete tokens in a more direct manner to
reveal the prosody encoding in discrete tokens.

\subsection{Evaluation of the Properties of Semantic Tokens}
Recent studies on the analysis of semantic token properties can be categorized into three directions:
(1) examining the correspondence between tokens and predefined labels such as speaker identity or phonemes\cite{shi2023, hubert},
(2) evaluating token robustness to non-linguistic information based on edit distance between tokens \cite{gat-etal-2023-augmentation, vashishth2024stabspeechtokenizerassessment}, and
(3) benchmarking the capabilities of speech language models built using discrete tokens \cite{deseyssel23_interspeech, seyssel-etal-2024-emphassess, maimon2025salmon}.
However, among these studies, very few have focused specifically on prosody. 
Though studies categorized under (3) tend to address prosody (pause\cite{deseyssel23_interspeech}, emphasis \cite{seyssel-etal-2024-emphassess}, and sentiment \cite{maimon2025salmon}),
these works serve as 
benchmarks for speech language models rather than for discrete tokens.
Thus, while existing work enables comparisons between pretrained language models,
there is a notable lack of research that investigates the design of discrete tokens themselves
which are used as the core of these models, from a prosodic perspective.
To fill in the gap, this study benchmarks discrete tokens trained under various conditions 
in terms of their ability to capture prosody.
In the following experiments, we utilize speech with artificially modified prosody 
to analyze how such changes are reflected in the tokens.

\section{Prosody and Speaker Modification}
\label{sec:mod}
In this study, we employ signal processing-based speech modification techniques 
to accurately investigate the prosody encoding capabilities of discrete tokens 
while clearly distinguishing between linguistic, paralinguistic, and non-linguistic information. 
By artificially modifying only prosody or only speaker timbre while keeping other elements, 
we analyze their effects on discrete tokens. %and downstream tasks.
\subsection{WORLD vocoder}
The WORLD vocoder \cite{morise2016world} is a high-quality speech analysis and synthesis system that
decomposes speech into three components: fundamental frequency ($f_0$), spectral envelope ($sp$), and aperiodicity.
Each of these components is represented as a frame-wise feature with a temporal dimension.
By manipulating these, we modify the prosody and
also the speaker identity (vocal tract length) of the speech signal.
This study focuses on pitch and intensity with respect to prosody. 
Since discrete tokens are frame-wise representations and deduplication is often applied to
ignore durational information, we did not focus on the duration in this study.
\subsection{Prosody Modification}
\subsubsection{Word-level Prosody Modification} \leavevmode\\
We sometimes emphasize a particular word by speaking it louder and 
with a higher pitch in order to convey our intention accurately. 
We simulate this by modifying the $f_0$ and $sp$
of the frames corresponding to the target word.\\
\noindent
\textbf{Pitch:} As shown in the following equation, 
the pitch is multiplied by $\alpha$ only for the frames (from $\text{start}_w$ to $\text{end}_w$)
which correspond to the target word $w$.
\begin{equation}
\label{eq:wordpitch}
    \tilde{f}_0^{(t)} = \begin{cases}
        \alpha \cdot f_0^{(t)} & \text{if } \text{start}_w \leq t < \text{end}_w \\
        f_0^{(t)} & \text{otherwise}
    \end{cases}
\end{equation}\\
\noindent
\textbf{Intensity:} While $sp$ has a frequency-axis dimension for each frame,
we can modify the intensity of the entire frame by multiplying it by $\beta$ equally along the frequency axis.
\begin{equation}
\label{eq:wordint}
    \tilde{sp}^{(f, t)} = \begin{cases}
        \beta \cdot sp^{(f, t)} & \text{if } \text{start}_w \leq t < \text{end}_w \\
        sp^{(f, t)} & \text{otherwise}
    \end{cases}
\end{equation}
where $f$ is the index of the frequency bin of the spectral envelope.

\subsubsection{Utterance-level Prosody Modification} \leavevmode\\
When we speak, not only the prosody of individual words but also the prosodic patterns of 
the entire utterance can convey implicit meanings.
In this study, we focus on the dynamism of prosody and modify prosodic features accordingly.\\
\noindent
\textbf{Pitch:} Focusing on pitch variation within an utterance, we adjust its range
by scaling the deviation from the mean pitch of the entire utterance $\mu_{\text{pitch}}$.
\begin{equation}
\label{eq:uttpitch}
\begin{aligned}
    \mu_{\text{pitch}} &= \frac{1}{|\mathcal{V}|} \sum_{t \in \mathcal{V}} f_0^{(t)}, \quad \mathcal{V} = \{ t \mid f_0^{(t)} > 0 \} \\
    \tilde{f}_0^{(t)} &=
    \begin{cases}
    \alpha \cdot (f_0^{(t)} - \mu_{\text{pitch}}) + \mu_{\text{pitch}} & \text{if } t \in \mathcal{V} \\
    f_0^{(t)} & \text{otherwise}
    \end{cases}
\end{aligned}
\end{equation}
where $\mathcal{V}$ is the set of frames where $f_0$ is not 0 (i.e., voiced frames).\\
\noindent
\textbf{Intensity:} Similar to the pitch, we can modify the dynamism of the intensity
by modifyng $sp$. We conduct this process in a logarithmic space to avoid
excessive amplification of the intensity. We first calculate the intensity of each frame $\text{int}^{(t)}$
by averaging the spectral envelope $sp$ across the frequency axis.
Then, we calculate the mean of the logarithm of the intensity $\mu_{\log(\text{int})}$
and scale the intensity of the voiced frames by $\beta$.
For each frame, $sp$ is scaled by the ratio of the modified intensity $\tilde{\text{int}}^{(t)}$ to the original intensity $\text{int}^{(t)}$,
uniformly across all frequency bins.
\begin{equation}
\label{eq:uttint}
\resizebox{0.91\linewidth}{!}{
$\begin{aligned}
    \text{int}^{(t)} &= \frac{1}{F} \sum_{f=1}^{F} sp^{(f, t)} \\
    \mu_{\log(\text{int})} &= \frac{1}{|\mathcal{V}|} \sum_{t \in \mathcal{V}}  \log(\text{int}^{(t)}), \quad \mathcal{V} = \{ t \mid f_0^{(t)} > 0 \} \\
    \tilde{\log(\text{int})}^{(t)} &= \begin{cases}
        \beta \cdot (\log(\text{int}^{(t)}) - \mu_{\log(\text{int})}) + \mu_{\log(\text{int})} & \text{if } t \in \mathcal{V} \\
        \log(\text{int}^{(t)}) & \text{otherwise}
    \end{cases} \\
    \tilde{sp}^{(f, t)} &= \exp{\left(\tilde{\log(\text{int})}^{(t)} - \log(\text{int}^{(t)})\right)} \cdot sp^{(f, t)} \\ 
\end{aligned}$
}
\end{equation}
where $F$ denotes the number of frequency bins.

\subsection{Speaker Modification}
To compare between speakers, it is possible to use utterances from different speakers 
(as we try in the following experiments); however, their prosodic features 
are not necessarily identical. Therefore, to strictly separate and compare speaker information 
and prosodic information, we generate speech in which only the speaker characteristics 
are modified by converting the $sp$.
By stretching the $sp$ along the frequency axis with $\gamma$, 
we simulate changes in the vocal tract length and modify the speaker identity.
\begin{equation}
\label{eq:spk}
\resizebox{0.91\linewidth}{!}{
$\begin{aligned}
f' &= \frac{f}{\gamma} \\
\tilde{sp}^{(f, t)} &= \left(1 + \lfloor f'\rfloor - f'\right) \cdot sp^{\left(\lfloor f'\rfloor, t\right)} + \left( f' - \lfloor f'\rfloor \right) \cdot sp^{\left(\lfloor f'\rfloor + 1, t\right)}
\end{aligned}$
}
\end{equation}
where $f'$ is the corresponding continuous position on the original frequency axis, 
and linear interpolation is applied to obtain the modified $sp$.

\section{Tokenization Strategy}
\label{sec:tokenize}
When obtaining discrete tokens by applying k-means clustering to the outputs of SSL models, various heuristic choices must be made, including the selection of the SSL model, the number of clusters, and other preprocessing steps. In this study, we investigate the following four aspects.
\subsection{SSL models}
\label{subseq:ssl}
We compare the following four SSL models, each trained with different objectives.
\\
\noindent
\textbf{HuBERT:} HuBERT\cite{hubert} is used to obtain discrete tokens in speech language models such as GSLM and its variants \cite{lakhotiaetal2021generative, kharitonov-etal-2022-text, nguyen-etal-2023-generative}. It is trained via masked prediction, where the objective is to predict labels obtained by discretizing acoustic features. From the second iteration onward, the model uses its own outputs instead of acoustic features.
\\
\noindent
\textbf{ContentVec:} ContentVec\cite{qian2022contentvec} can be regarded as a speaker-invariant extension of HuBERT. Using a pretrained HuBERT as the teacher, the student model is trained to predict the teacher's labels on training data where speaker identity has been unified via voice conversion. A contrastive loss is additionally employed to enforce invariance to formant frequency and f0 scaling. 
\\
\noindent
\textbf{data2vec:} data2vec\cite{baevski2022data2vec} is a model trained through a self-distillation approach, in which the model learns to predict continuous feature representations of masked inputs. In \cite{ma2024icassp}, data2vec has been shown to achieve strong performance in speech emotion recognition when used as continuous features. In this study, we investigate its prosodic encoding capability when the data2vec features are discretized.
\\
\noindent
\textbf{emotion2vec:} emotion2vec\cite{ma-etal-2024-emotion2vec} is an emotion-aware version of data2vec. In emotion2vec, an additional loss is introduced to predict utterance-level emotional representations, in addition to the frame-level masked prediction. This enables the model to capture both fine-grained and global emotional characteristics within speech. The model is trained on emotional speech data, further enhancing its ability to capture affective information.

For all models, we used the ``base" architecture consisting of 12 transformer layers. For HuBERT\footnote{\url{https://huggingface.co/facebook/hubert-base-ls960}}, ContentVec\footnote{legacy\_500 in \url{https://github.com/auspicious3000/contentvec}}, and data2vec\footnote{\url{https://huggingface.co/facebook/data2vec-audio-base}}, we used the ones trained on LibriSpeech960 \cite{libri}. For emotion2vec\footnote{\url{https://huggingface.co/emotion2vec/emotion2vec_base}}, we used the one that was initialized with a data2vec model pretrained on LibriSpeech960, and subsequently trained on 262 hours of unlabeled emotional speech data.

\subsection{Cluster sizes}
\label{subseq:size}
In previous studies, a variety of cluster sizes have been used on k-means clustering. We also compare multiple cluster sizes to examine how the number of clusters affects the prosody encoding. Specifically, we conduct experiments with cluster sizes of 100, 500, and 2000.

\subsection{Training data for k-means clustering}
\label{subseq:mead}
In \cite{onda2025discretetokensexhibitinterlanguage}, it has been revealed that in discrete token-based ASR, recognition accuracy for foreign-accented speech can be improved by training the k-means clustering on the speaker’s native language, even when the upstream SSL model remains the same. This highlights that the training data used for k-means clustering is also an important factor influencing the nature of the resulting tokens. While prior studies have typically used neutral read speech such as LibriSpeech for learning cluster centroids, this study investigates whether using emotional speech to train the k-means model enhances sensitivity to prosodic information. Here, we utilized MEAD \cite{mead}, a TIMIT-based relatively large emotional speech dataset. We compared the results of k-means clustering trained on MEAD and LibriSpeech.

\subsection{Moving average of SSL features} %(\textbf{\textit{ours}})}
\label{subseq:ma}
In \cite{kando2025exploringeffectsegmentationvocabulary}, a preprocessing method was explored in which the SSL feature sequence was segmented into fixed-length frame windows, and the features within each window were averaged into a single vector. In this study, we extend this approach by applying a moving average to the SSL features $\mathbf{z}_i (0\leq i \leq T-1)$ with a window size $W=2w+1$, as follows.
\begin{equation}
\begin{aligned}
    t_{i,\text{start}}, t_{i,\text{end}} &= \max(0, i - w), \min(T-1, i + w) \\
    \bar{\mathbf{z}}_i &= \frac{1}{t_{i,\text{end}} - t_{i,\text{start}} + 1}\sum_{j=t_{i,\text{start}}}^{t_{i,\text{end}}} \mathbf{z}_j 
\end{aligned}
\end{equation}
It is known that humans perceive prosody at a temporal resolution longer than phonetic segments \cite{POEPPEL2003245}. This motivates our approach, which aims to capture prosodic variation over longer time spans while preserving the continuous nature of speech.

% \subsection{Voting over discrete tokens (\textbf{\textit{ours}})}
% K-means clustering may assign subtle differences in the continuous feature space to different clusters, potentially resulting in noisy discrete token sequences. One straightforward approach to mitigate this issue is to apply majority voting within a sliding window over the discrete tokens, thereby smoothing the output.

\section{Experiments}
%%%%%%%%%%%%%%%
\begin{figure*}[t]
  \centering
  \hspace*{-5mm}
  \includegraphics[width=1.02\textwidth]{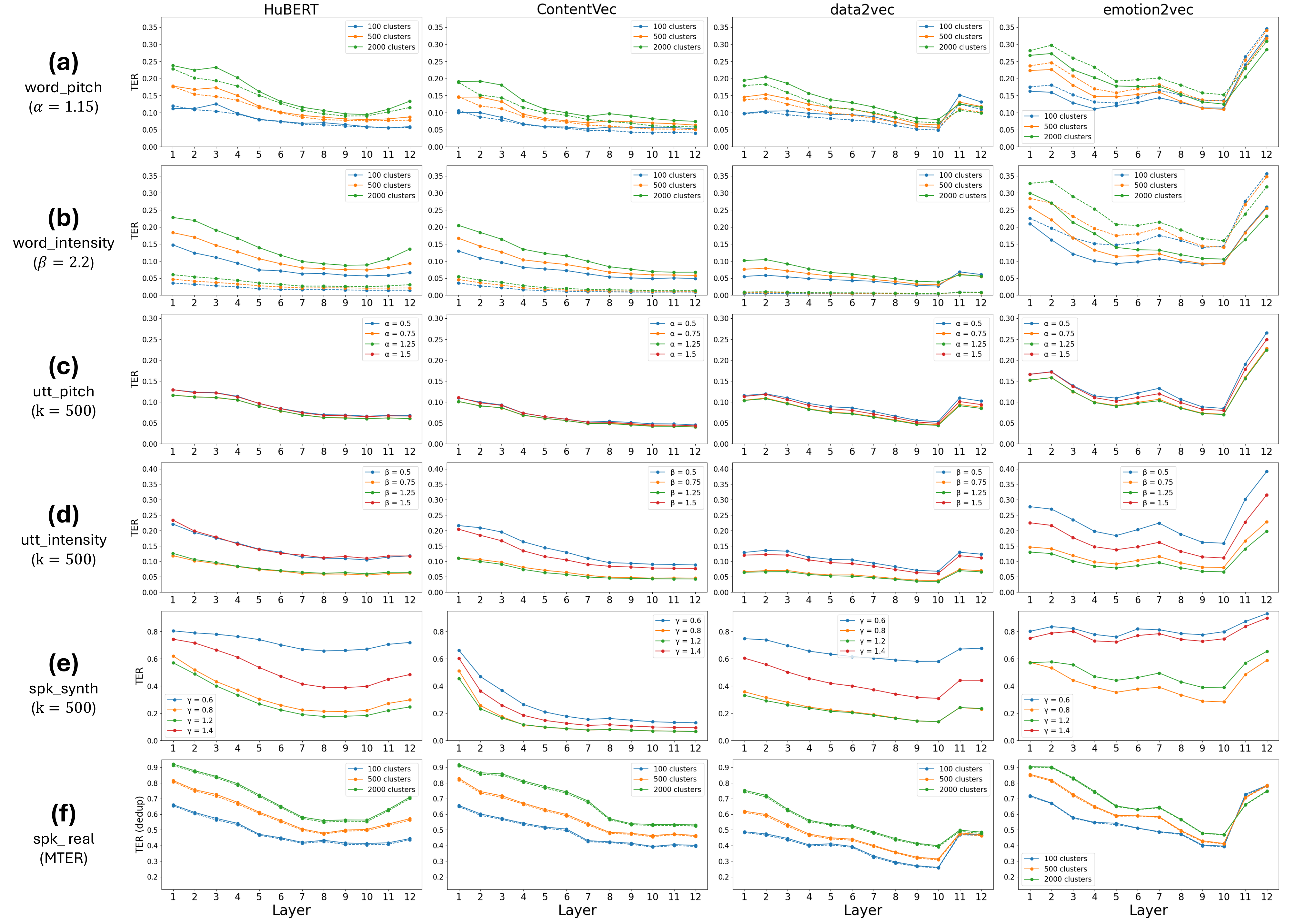}
  \vspace*{-5mm}
  \caption{Comparison of Token Error Rate (TER) between modified and original (resynthesized) speech across SSL models:
  \textbf{(a)(b)} TER on target words when the pitch or intensity of a specific word is modified as described in Eq.(\ref{eq:wordpitch}) and (\ref{eq:wordint}). Dotted lines indicate TER when the entire utterance is modified at the same level.
  \textbf{(c)(d)} TER on entire utterances when the pitch and intensity is modified at the utterance level, as described in Eq.(\ref{eq:uttpitch}) and (\ref{eq:uttint}). This figure shows the case where the number of clusters is 500.
  \textbf{(e)} TER when the speaker identity is modified following Eq.(\ref{eq:spk}). The number of clusters is 500.
  \textbf{(f)} TER between real utterances from different speakers with the same sentence. Unlike (a)--(e), since there is no definitive ground-truth reference, we compute the average TER over all possible pairs (MTER). Since the utterance lengths may vary across speakers, we apply deduplication before computing TER.}
  \label{fig:res1}
  \vspace*{-2mm}
\end{figure*}
%%%%%%%%%%%%%%%%%%
We applied the speech modification method described in Section \ref{sec:mod} to the TIMIT corpus \cite{timit} to create the evaluation data. The resulting speech was then tokenized using various discretization methods described in Section \ref{sec:tokenize}, and we analyzed how the discrete token sequences changed before and after the modification on prosody and speaker identity.
TIMIT provides manually annotated alignments, enabling more precise word-level speech modifications. In addition, since the same sentences are read by multiple speakers, it allows for comparisons not only between artificially modified speech but also between real different human utterances. In this study, we used the SX set.
For tokenization, unless explicitly specified, k-means clustering was trained on the 30-hour random subset of LibriSpeech-100h \cite{libri}, with no preprocessing applied to the SSL features and no postprocessing to the resulting discrete token sequences.
\subsection{SSL models and cluster sizes}
\label{subsec:sslresult}
\begin{figure}[t]
  \centering
  \includegraphics[width=0.9\columnwidth]{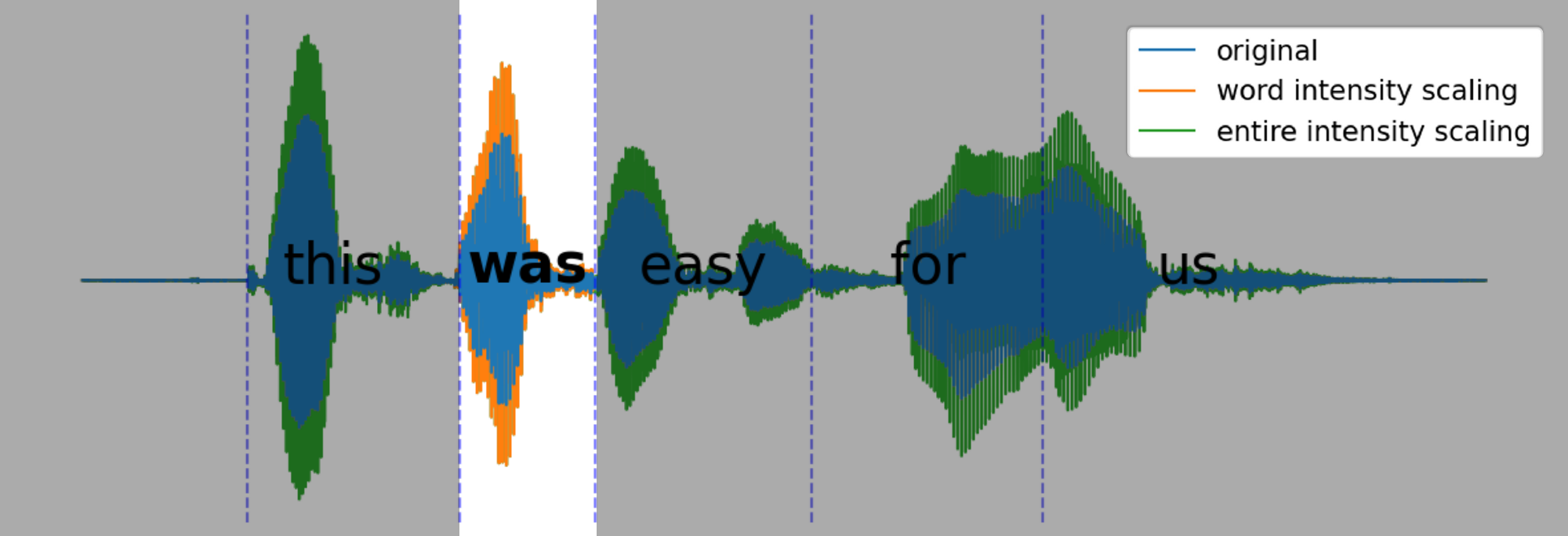}
  \caption{Overview of word-level prosody comparison: scaling on specific words (orange) and entire utterance (green). Token sequences are compared only within the target word segment.}
  \label{fig:word}
  \vspace*{-3mm}
\end{figure}

We conduct a comprehensive comparison of the four SSL models described in Section \ref{subseq:ssl}, focusing on their encoding of prosody and speaker information. We also examine the effects of varying the number of clusters (as discussed in \ref{subseq:size}) and the choice of transformer layers. 
The evaluation was conducted by computing the Token Error Rate (TER), i.e., the edit distance between the discrete token sequences of the original (analysis-resynthesis) speech and those of the modified speech. 
% As the WORLD vocoder preserves timing information, the sequence lengths remain exactly the same in both cases.
Fig. \ref{fig:res1} presents the overall results. In the following, we highlight notable findings and discuss the insights derived from them.
\vspace*{-3mm}
\\\\
\noindent
\textbf{Word-level prosody modification:} 
Following Eqs. (\ref{eq:wordpitch}) and (\ref{eq:wordint}), we generated modified versions of each utterance in which only a single specific word was emphasized. We then computed TER between the token sequences for original and modified speech (blue vs. orange in Fig. \ref{fig:word}). For each utterance, we repeated this procedure for all words, and averaged the TERs. To examine whether SSL models are sensitive to local or global prosodic changes, we also compared this with the case where the entire utterance was emphasized by the same scaling factor (blue vs. green in Fig. 2). In both cases, TER was computed only within the segment corresponding to the target word. As such, if we extract only the relevant segment from both versions, the resulting audio should be identical.

The results are shown in rows (a) and (b) of Fig. \ref{fig:res1}. $\alpha$ and $\beta$ in Eqs.(\ref{eq:wordpitch}) and (\ref{eq:wordint}) are set to 1.15 and 2.2, respectively. Solid lines show scaling of specific words; dashed lines show scaling of the entire utterance.
Across all conditions, HuBERT, ContentVec, and data2vec consistently exhibited higher sensitivity to local emphasis on specific words compared to global emphasis, both for pitch and intensity (the solid lines always lie above the dashed lines). This suggests that these models capture relative prosodic contours within an utterance, rather than absolute acoustic quantities at each time frame. This tendency is especially strong for intensity; in all these three models, scaling the intensity of the entire utterance results in almost no change in the token sequence. This is likely because these models employ CNN encoders that take raw waveform as inputs. Compared to pitch, intensity directly affects the amplitude values of the waveform, and thus, when internal normalization is applied, differences in global intensity tend to be canceled out. It is explicitly stated in the wav2vec 2.0 paper \cite{wav2vec} that layer normalization is performed after the CNN encoder before feeding into the transformer blocks. Therefore, these models, that adopt a similar architecture to wav2vec 2.0, are also considered insensitive to global scaling of utterance intensity. Additionally, it is stated in the data2vec paper that the input waveform is normalized to zero mean and unit variance before being fed into the CNN encoder, which likely further reduces the sensitivity compared to the other two models. Unlike these three models, emotion2vec exhibited greater sensitivity to global prosody scaling for both pitch and intensity. This is likely because, in addition to the frame-wise masked prediction loss, emotion2vec introduces an utterance-level loss based on representations obtained through temporal mean pooling. Here, we may say that frame-level masked prediction is effective in capturing relative variations within an utterance, rather than absolute physical quantities.

\begin{figure}[t]
  \centering
  \includegraphics[width=\columnwidth]{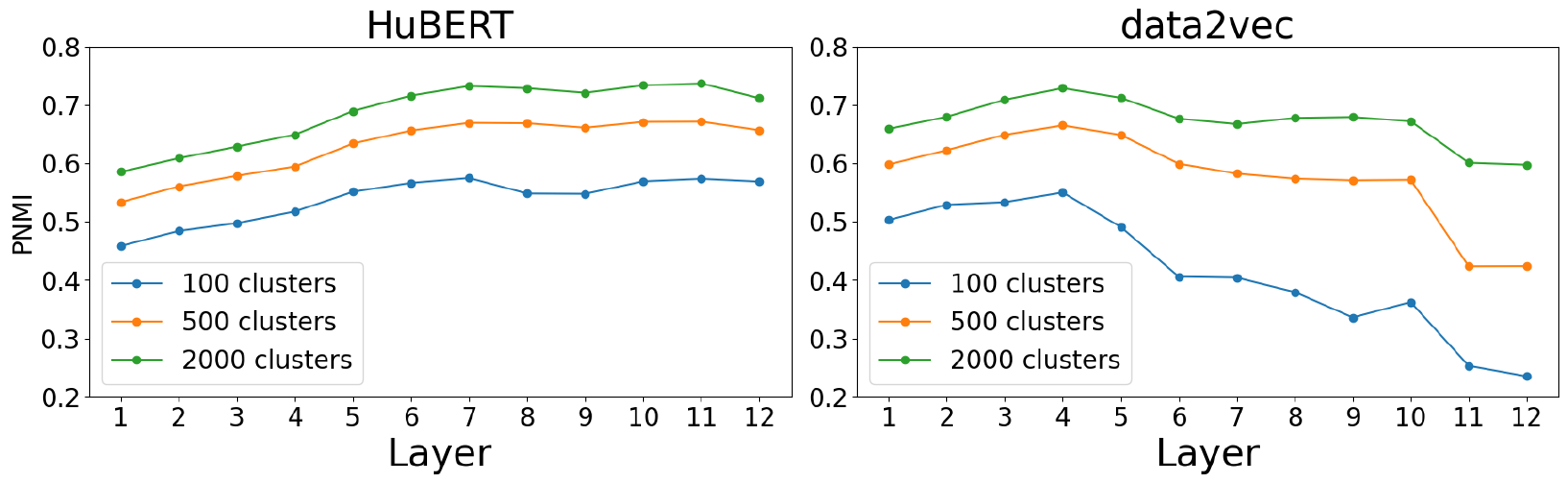}
  \vspace*{-6mm}
  \caption{Comparison of phone normalized mutual information (PNMI): HuBERT vs. data2vec. Higher values indicate better alignment with phoneme.}
  \label{fig:pnmi}
  \vspace*{-3mm}
\end{figure}

Regarding cluster sizes, larger numbers generally led to increased sensitivity to prosodic variations; however, in layers 11 and 12 of data2vec and emotion2vec, smaller clusters exhibited higher sensitivity. 
This is likely because both models are trained using masked prediction targeting continuous features, thus their representations are not inherently ``clustering-friendly."
In contrast, HuBERT and Contentvec are trained to predict discretized token labels. HuBERT is trained in an iterative manner, and it is known that the discrete targets become increasingly aligned with phonemes \cite{hubert}. As a result, the resulting continuous feature space is also expected to conform more closely to phoneme-based categorization.
Fig.\ref{fig:pnmi} presents a comparison of phone normalized mutual information (PNMI), a commonly used metric for evaluating the alignment between discrete tokens and phonemes \cite{hubert, chang23b_interspeech, onda2025differentiablekmeansfullyoptimizeddiscrete}. Here, PNMI is calculated using only original speech data from TIMIT without any modifications. It can be observed that for HuBERT, PNMI increases in the later layers regardless of the number of clusters. In contrast, data2vec shows a sharp drop in PNMI after the 10th layer, especially when using smaller cluster sizes, indicating that phonemic distinctions are not well captured in such settings.
When treated as continuous features, data2vec is known to exhibit superior ASR performance than HuBERT, suggesting that it contains sufficient linguistic information \cite{baevski2022data2vec, yoon23c_interspeech}. Indeed, as shown in Fig. \ref{fig:res1} (a) (b), though data2vec demonstrates inferior robustness to prosodic variations than HuBERT at 100 clusters, as the cluster size increases and the discrete tokens approaches a continuous representation (i.e., at 2000 clusters), data2vec shows lower sensitivity to such variations than HuBERT. These results suggest that when using a small number of clusters, SSL models trained to classify discrete labels are more suitable, whereas with larger cluster sizes, closer to continuous representations, models trained to reconstruct continuous targets perform better. In other words, while masked prediction enables the acquisition of linguistic information using both discrete and continuous targets, using discrete representations is more effective for learning phoneme-like segmental representations. However, these trends are prominent only near the last layer (i.e., 11th and 12th layers), while all models exhibit relatively stable linguistic representations around layers 9 and 10.
\vspace*{-3mm}
\\\\
\noindent
\textbf{Utterance-level prosody modification:} Following Eqs. (\ref{eq:uttpitch}) and (\ref{eq:uttint}), we simulated variations in emotional intensity. The results are presented in Fig. \ref{fig:res1} (c) and (d). These results largely align with those observed in (a) and (b), where SSL models generally capture the overall prosodic contour. Among them, emotion2vec demonstrated particularly high sensitivity.
\vspace*{-3mm}
\\\\
\noindent
\textbf{Speaker variations:} To investigate the speaker-invariance of discrete tokens, we used pseudo speaker-modified utterances generated as in Eq.(\ref{eq:spk}) (row (e) in Fig.\ref{fig:res1}), along with actual recordings of the same sentences spoken by different speakers in the TIMIT corpus (shown in row (f)). For the latter, since sequence lengths vary, we performed deduplication before computing TER. In addition, since no absolute ground truth can be defined, we calculated the mean token error rate (MTER)\cite{onda2025differentiablekmeansfullyoptimizeddiscrete} by averaging TER across all possible speaker pairs. In row (e), ContentVec shows a notably low value, which can likely be attributed to the use of a contrastive loss against formant shifts during training. However, in row (f), it performs similarly to HuBERT in all but the last two layers, suggesting that its speaker-invariance in response to actual speaker differences remains limited. In contrast, data2vec demonstrates clearly higher speaker robustness in both (e) and (f), indicating that discretization of the masked prediction target is not necessary for abstracting away speaker identity. An interesting finding is observed with emotion2vec: in row (f), for example at layer 10 with 2000 clusters, it shows relatively high speaker invariance (lower value in TER), data2vec (0.40) $<$ emotion2vec (0.47) $<$ contentvec (0.54) $<$ HuBERT (0.56). Given emotion2vec’s high sensitivity to prosodic changes as seen in rows (a)–(d), this suggests that the discrete tokens derived from emotion2vec possess an ideal property for characterizing prosody in a speaker-invariant manner.

In (f), the dotted lines indicate results for speaker pairs sharing the same regional accent. The minimal difference compared to results that included cross-accent pairs (shown in solid lines) suggests that these SSL models exhibit robustness to subtle pronunciation variations.

\subsection{Training data for k-means clustering}
\begin{table}[tb]
    \centering
    \caption{Comparison based on training data for k-means clustering: Librispeech vs. MEAD. (a) and (f) correspond to those in Fig. \ref{fig:res1}. \textbf{Bold} indicates a significantly higher score ($p < 0.05$).}
    \label{tab:mead}
    \vspace*{-1mm}
    \resizebox{\columnwidth}{!}{
        \begin{tabular}{ccccccccccc}
	  \toprule
	   &  &  & \multicolumn{2}{c}{HuBERT} & \multicolumn{2}{c}{ContentVec} & \multicolumn{2}{c}{data2vec} & \multicolumn{2}{c}{emotion2vec}\\
        metric & layer & k & Libri & MEAD & Libri & MEAD & Libri & MEAD & Libri & MEAD\\
	\midrule
        \multirow{2}{*}{\Large\textbf{(a)}} & 10 & 100 & 0.059& \textbf{0.071}& 0.056& \textbf{0.059}& 0.056& 0.057& 0.113& \textbf{0.122}\\
               &    &2000 & 0.094& \textbf{0.110}& 0.082& \textbf{0.088}& 0.080& \textbf{0.086}& 0.125& \textbf{0.135} \\\cdashline{2-11}\noalign{\vskip 2pt}
        word   & 12 & 100 & 0.059& \textbf{0.068}& 0.052& \textbf{0.057}& \textbf{0.131}& 0.128& \textbf{0.324}& 0.317\\
        pitch  &    &2000 & 0.134& 0.134& 0.074& \textbf{0.083}& 0.116& \textbf{0.125}& 0.285& \textbf{0.320}\\\midrule
        \multirow{2}{*}{\Large\textbf{(f)}} & 10 & 100 & 0.413& \textbf{0.473}& 0.395& \textbf{0.413}& 0.260& \textbf{0.269}& 0.397& \textbf{0.470}\\
               &    &2000 & 0.563& \textbf{0.616}& 0.535& \textbf{0.560}& 0.399& \textbf{0.414}& 0.470& \textbf{0.538} \\\cdashline{2-11}\noalign{\vskip 2pt}
         spk   & 12 & 100 & 0.444& \textbf{0.479}& 0.401& \textbf{0.416}& \textbf{0.467}& 0.454& \textbf{0.785}& 0.750\\
        real &  &2000 & 0.710& \textbf{0.718}& 0.532& \textbf{0.574}& 0.486& \textbf{0.512}& 0.750& \textbf{0.801}\\%\midrule
        % \multirow{2}{*}{\textbf{PNMI}}& 10 & 100 & \textbf{0.569}& 0.393& \textbf{0.421}& 0.337& \textbf{0.362}& 0.330& \textbf{0.263}& 0.228\\
        %         &    &2000 & \textbf{0.733}& 0.715& 0.683& \textbf{0.712}& 0.671& \textbf{0.733}& 0.605& \textbf{0.657}\\\cdashline{2-11}\noalign{\vskip 2pt}
        %         & 12 & 100 & \textbf{0.569}& 0.397& \textbf{0.491}& 0.376& \textbf{0.235}& 0.200& 0.132& \textbf{0.179}\\
        %         &    &2000 & 0.712& \textbf{0.715}& 0.727& \textbf{0.728}& 0.598& \textbf{0.667}& 0.443& \textbf{0.527}\\
	  \bottomrule
	\end{tabular}
    }
  \vspace*{-1mm}
\end{table}
\begin{figure}[t]
  \centering
  \includegraphics[width=\columnwidth]{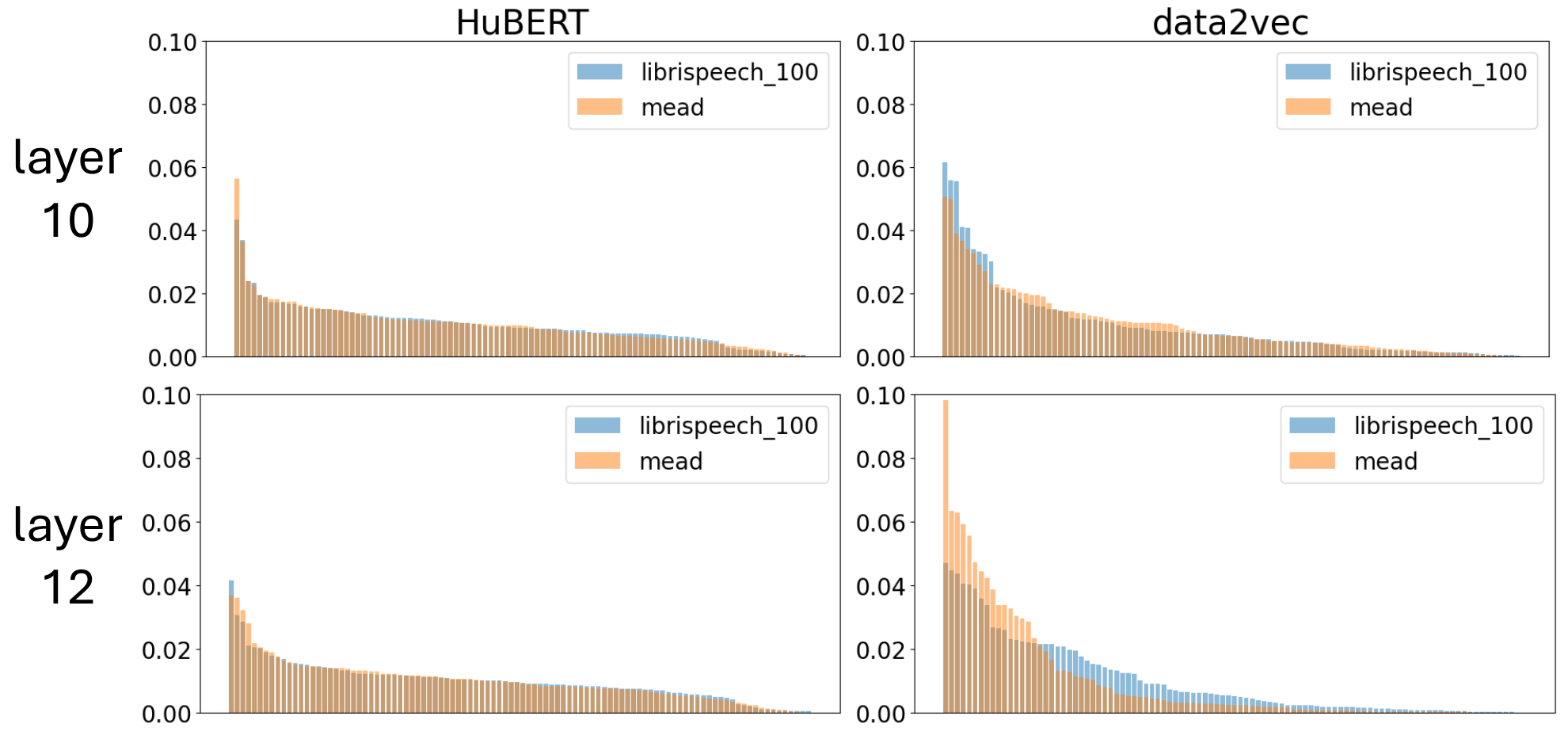}
  \vspace*{-6mm}
  \caption{Sorted normalized frequency distributions of clusters: the 10th and 12th layers of HuBERT and data2vec ($k=100$)}
  \label{fig:distribution}
  \vspace*{-3mm}
\end{figure}
As discussed in \ref{subseq:mead}, we investigate whether using emotional speech to train the k-means clustering increases the sensitivity of the discrete tokens to prosodic variation. We compared LibriSpeech-100h (neutral read speech)\cite{libri} and MEAD (emotional read speech)\cite{mead}, using 30-hour random subsets from each dataset.

The results are shown in Table \ref{tab:mead}. As a measure of prosodic sensitivity, we report only word-level pitch modification (corresponding to Fig. \ref{fig:res1} (a)), as the results for other prosodic manipulations (Fig. \ref{fig:res1} (b)–(d)) showed mostly same trends. In addition, speaker invariance under actual speaker variation (Fig. \ref{fig:res1} (f)) are also shown. A paired t-test was conducted to compare the performance between LibriSpeech and MEAD. Values that were significantly higher at the $p<0.05$ level are indicated in bold.
In most cases, training the k-means clustering model on MEAD led to improved sensitivity to prosody and speaker variation. This is presumably because the clusters are learned to reflect not only phonetic features but also prosodic characteristics at the fine-grained cluster boundaries. An opposite trend was observed when using the last layer of data2vec and emotion2vec with cluster size of 100. This is likely due to the ``clustering-unfriendly" nature of the representations from these two models (as discussed in Section \ref{subsec:sslresult}). 
Fig. \ref{fig:distribution} shows the frequency distributions of clusters for the TIMIT dataset, when k-means clustering was trained using the 10th and 12th (final) layer of HuBERT and data2vec, with the cluster size of 100. For HuBERT, the cluster frequency distributions remain nearly identical regardless of whether the k-means clustering was trained on neutral or emotional speech, for both 10th and 12th layers. This suggests that the clustering robustly captures phoneme-based structures, largely unaffected by the prosodic richness of the training data. In contrast, the 12th layer of data2vec exhibits substantial shifts in cluster distributions depending on the training corpus. The distribution learned from MEAD shows a strong imbalance. The more skewed distribution likely led to assigning speech segments into the same clusters more frequently, thereby reducing the sensitivity to prosodic variations. On the other hand, when using the 10th layer, the same trend as HuBERT was observed even in data2vec. This indicates that the differences in the characteristics of the representations resulting from the pretraining objectives (i.e., discrete labels or continuous representations) are mainly reflected in the final layers, whereas the intermediate layers exhibit largely similar capabilities of the encoding of linguistic information. A similar trend about model-specific patterns can also be seen in the speaker invariance of ContentVec: as shown in Fig. \ref{fig:res1}(f), the differences between HuBERT and ContentVec emerge primarily in layers 11 and 12.

\subsection{Moving average}
\begin{figure}[t]
  \centering
  \includegraphics[width=\columnwidth]{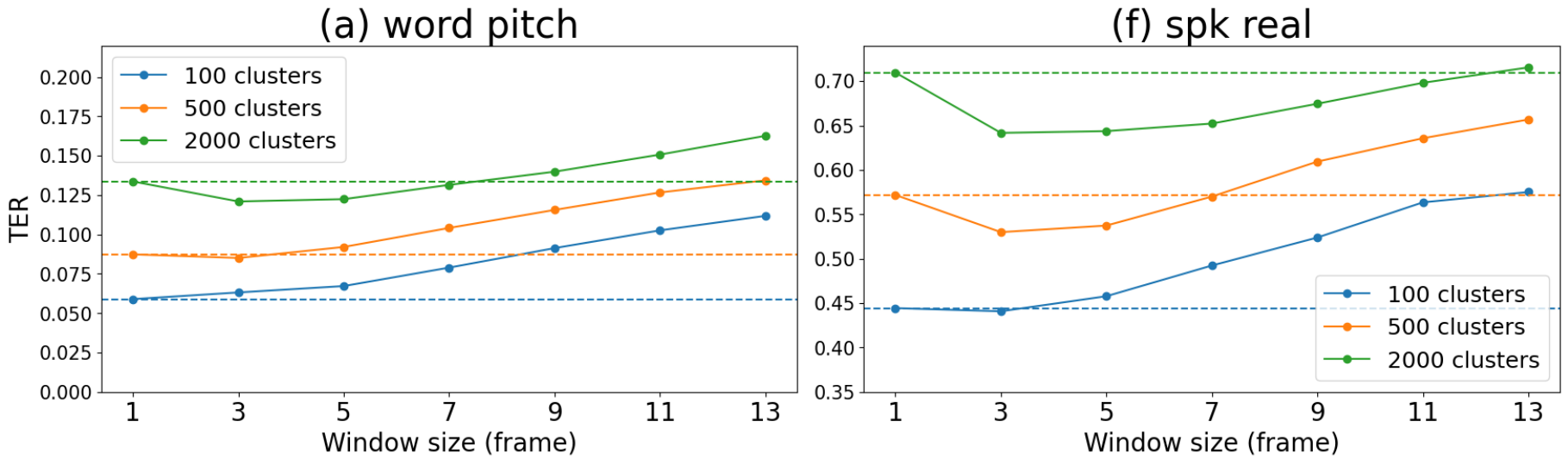}
  \vspace*{-6mm}
  \caption{Effect of applying moving average to SSL features on sensitivity to prosody and speaker variation (HuBERT 12th layer)}
  \label{fig:ma}
  \vspace*{-3mm}
\end{figure}

As discussed in Section \ref{subseq:ma}, we investigate the effect of applying a moving average to SSL features as a preprocessing step. Fig. \ref{fig:ma} shows the results of comparing sensitivity to prosodic and speaker variation after smoothing with different window sizes for the final layer of HuBERT. For both prosody and speaker variation, increasing the window size for the moving average initially leads to a decrease in sensitivity, followed by a gradual increase. However, this increase is more moderate for speaker variation. For instance, with 2000 clusters, applying a moving average with a window size of 9 or 11 results in higher sensitivity to prosody compared to the no-smoothing condition (window size = 1), while reducing sensitivity to speaker. This suggests that, by selecting an appropriate window size for the moving average according to the cluster size, it is possible to design discrete tokens that better capture desired characteristics such as prosodic sensitivity or speaker invariance. This can be attributed to the differing temporal resolutions required to capture each characteristic.
% ; in an extreme case, speaker identity may be sufficiently represented by averaging over the entire utterance.

\section{Conclusions}
Through a comprehensive set of analytical experiments, we demonstrated that the sensitivity of discrete tokens to prosody and speaker identities can be effectively controlled by appropriately combining various factors involved in discretization. 
In particular, the following findings could be useful guidance for better design of discrete tokens: 
1) SSL models trained with frame-wise masked prediction tend to capture relative prosodic contours well, whereas models that promote utterance-level consistency (i.e., emotion2vec) also exhibit sensitivity to absolute prosodic variations across the entire utterance,
2) SSL models pretrained with discrete label prediction capture linguistic information well even with small cluster sizes, while models trained with continuous value prediction become effective with larger cluster sizes (these differences, however, were mainly observed in the final layers, whereas the intermediate layers exhibited largely similar behavior.), 
3) training k-means clustering with emotional speech increases sensitivity to prosody, and 
4) applying a moving average to SSL features with an appropriate window size can improve both speaker invariance and prosodic sensitivity.

A limitation of this study is that the analysis is primarily based on the tokens’ sensitivity to input variations. Future work should investigate whether linguistic, prosodic, and speaker information can actually be extracted from the resulting tokens.
% \section*{Acknowledgements}
% \vspace*{1mm}
\\\\
\noindent
\textbf{Acknowledgements} \hspace{1mm} This work was supported by AIST policy-based budget project ``R\&D on Generative AI Foundation Models for the Physical Domain"

\bibliographystyle{IEEEtran}
\bibliography{references}

% Generated by IEEEtran.bst, version: 1.12 (2007/01/11)
\begin{thebibliography}{10}
\providecommand{\url}[1]{#1}
\csname url@samestyle\endcsname
\providecommand{\newblock}{\relax}
\providecommand{\bibinfo}[2]{#2}
\providecommand{\BIBentrySTDinterwordspacing}{\spaceskip=0pt\relax}
\providecommand{\BIBentryALTinterwordstretchfactor}{4}
\providecommand{\BIBentryALTinterwordspacing}{\spaceskip=\fontdimen2\font plus
\BIBentryALTinterwordstretchfactor\fontdimen3\font minus \fontdimen4\font\relax}
\providecommand{\BIBforeignlanguage}[2]{{%
\expandafter\ifx\csname l@#1\endcsname\relax
\typeout{** WARNING: IEEEtran.bst: No hyphenation pattern has been}%
\typeout{** loaded for the language `#1'. Using the pattern for}%
\typeout{** the default language instead.}%
\else
\language=\csname l@#1\endcsname
\fi
#2}}
\providecommand{\BIBdecl}{\relax}
\BIBdecl

\bibitem{hubert}
W.-N. Hsu, B.~Bolte, Y.-H.~H. Tsai, K.~Lakhotia, R.~Salakhutdinov, and A.~Mohamed, ``Hu{BERT}: Self-supervised speech representation learning by masked prediction of hidden units,'' \emph{T-ASLP}, vol.~29, pp. 3451--3460, 2021.

\bibitem{wav2vec}
A.~Baevski, Y.~Zhou, A.~Mohamed, and M.~Auli, ``wav2vec 2.0: A framework for self-supervised learning of speech representations,'' in \emph{NeurIPS 2020}, vol.~33, 2020, pp. 12\,449--12\,460.

\bibitem{Chen2021WavLMLS}
\BIBentryALTinterwordspacing
S.~Chen, C.~Wang, Z.~Chen, Y.~Wu, S.~Liu, Z.~Chen, J.~Li, N.~Kanda, T.~Yoshioka, X.~Xiao, J.~Wu, L.~Zhou, S.~Ren, Y.~Qian, Y.~Qian, M.~Zeng, and F.~Wei, ``Wav{LM}: Large-scale self-supervised pre-training for full stack speech processing,'' \emph{IEEE Journal of Selected Topics in Signal Processing}, vol.~16, pp. 1505--1518, 2021. [Online]. Available: \url{https://api.semanticscholar.org/CorpusID:239885872}
\BIBentrySTDinterwordspacing

\bibitem{Chung2021w2vBERTCC}
\BIBentryALTinterwordspacing
Y.-A. Chung, Y.~Zhang, W.~Han, C.-C. Chiu, J.~Qin, R.~Pang, and Y.~Wu, ``w2v-{BERT}: Combining contrastive learning and masked language modeling for self-supervised speech pre-training,'' \emph{2021 IEEE Automatic Speech Recognition and Understanding Workshop (ASRU)}, pp. 244--250, 2021. [Online]. Available: \url{https://api.semanticscholar.org/CorpusID:237048255}
\BIBentrySTDinterwordspacing

\bibitem{sslreview}
A.~Mohamed, H.-y. Lee, L.~Borgholt, J.~D. Havtorn, J.~Edin, C.~Igel, K.~Kirchhoff, S.-W. Li, K.~Livescu, L.~Maal^^c3^^b8e, T.~N. Sainath, and S.~Watanabe, ``Self-supervised speech representation learning: A review,'' \emph{IEEE Journal of Selected Topics in Signal Processing}, vol.~16, no.~6, pp. 1179--1210, 2022.

\bibitem{guo2025recentadvancesdiscretespeech}
\BIBentryALTinterwordspacing
Y.~Guo, Z.~Li, H.~Wang, B.~Li, C.~Shao, H.~Zhang, C.~Du, X.~Chen, S.~Liu, and K.~Yu, ``Recent advances in discrete speech tokens: A review,'' 2025. [Online]. Available: \url{https://arxiv.org/abs/2502.06490}
\BIBentrySTDinterwordspacing

\bibitem{lakhotiaetal2021generative}
K.~Lakhotia, E.~Kharitonov, W.-N. Hsu, Y.~Adi, A.~Polyak, B.~Bolte, T.-A. Nguyen, J.~Copet, A.~Baevski, A.~Mohamed, and E.~Dupoux, ``On generative spoken language modeling from raw audio,'' \emph{TACL}, vol.~9, pp. 1336--1354, 2021.

\bibitem{zhang2023speechgpt}
D.~Zhang, S.~Li, X.~Zhang, J.~Zhan, P.~Wang, Y.~Zhou, and X.~Qiu, ``Speechgpt: Empowering large language models with intrinsic cross-modal conversational abilities,'' in \emph{Findings of the Association for Computational Linguistics: EMNLP 2023}, 2023, pp. 15\,757--15\,773.

\bibitem{borsos2023audiolm}
\BIBentryALTinterwordspacing
Z.~Borsos, R.~Marinier, D.~Vincent, E.~Kharitonov, O.~Pietquin, M.~Sharifi, D.~Roblek, O.~Teboul, D.~Grangier, M.~Tagliasacchi, and N.~Zeghidour, ``Audiolm: A language modeling approach to audio generation,'' \emph{IEEE/ACM Trans. Audio, Speech and Lang. Proc.}, vol.~31, p. 2523^^e2^^80^^932533, Jun. 2023. [Online]. Available: \url{https://doi.org/10.1109/TASLP.2023.3288409}
\BIBentrySTDinterwordspacing

\bibitem{rubenstein2023audiopalm}
P.~K. Rubenstein, C.~Asawaroengchai, D.~D. Nguyen, A.~Bapna, Z.~Borsos, F.~d.~C. Quitry, P.~Chen, D.~E. Badawy, W.~Han, E.~Kharitonov \emph{et~al.}, ``Audiopalm: A large language model that can speak and listen,'' \emph{arXiv preprint arXiv:2306.12925}, 2023.

\bibitem{arora2025landscapespokenlanguagemodels}
\BIBentryALTinterwordspacing
S.~Arora, K.-W. Chang, C.-M. Chien, Y.~Peng, H.~Wu, Y.~Adi, E.~Dupoux, H.-Y. Lee, K.~Livescu, and S.~Watanabe, ``On the landscape of spoken language models: A comprehensive survey,'' 2025. [Online]. Available: \url{https://arxiv.org/abs/2504.08528}
\BIBentrySTDinterwordspacing

\bibitem{chang24b_interspeech}
X.~Chang, J.~Shi, J.~Tian, Y.~Wu, Y.~Tang, Y.~Wu, S.~Watanabe, Y.~Adi, X.~Chen, and Q.~Jin, ``The {I}nterspeech 2024 challenge on speech processing using discrete units,'' in \emph{Interspeech 2024}, 2024, pp. 2559--2563.

\bibitem{chang23b_interspeech}
X.~Chang, B.~Yan, Y.~Fujita, T.~Maekaku, and S.~Watanabe, ``Exploration of efficient end-to-end {ASR} using discretized input from self-supervised learning,'' in \emph{Interspeech 2023}, 2023, pp. 1399--1403.

\bibitem{chang2024exploring}
X.~Chang, B.~Yan, K.~Choi, J.-W. Jung, Y.~Lu, S.~Maiti, R.~Sharma, J.~Shi, J.~Tian, S.~Watanabe, Y.~Fujita, T.~Maekaku, P.~Guo, Y.-F. Cheng, P.~Denisov, K.~Saijo, and H.-H. Wang, ``Exploring speech recognition, translation, and understanding with discrete speech units: A comparative study,'' in \emph{ICASSP 2024}, 2024, pp. 11\,481--11\,485.

\bibitem{yang2024towards}
Y.~Yang, F.~Shen, C.~Du, Z.~Ma, K.~Yu, D.~Povey, and X.~Chen, ``Towards universal speech discrete tokens: A case study for {ASR} and {TTS},'' in \emph{ICASSP 2024}, 2024, pp. 10\,401--10\,405.

\bibitem{mousavi24_interspeech}
P.~Mousavi, J.~Duret, S.~Zaiem, L.~{Della Libera}, A.~Ploujnikov, C.~Subakan, and M.~Ravanelli, ``How should we extract discrete audio tokens from self-supervised models?'' in \emph{Interspeech 2024}, 2024, pp. 2554--2558.

\bibitem{shi24h_interspeech}
J.~Shi, X.~Ma, H.~Inaguma, A.~Sun, and S.~Watanabe, ``{MMM}: Multi-layer multi-residual multi-stream discrete speech representation from self-supervised learning model,'' in \emph{Interspeech 2024}, 2024, pp. 2569--2573.

\bibitem{shen2024}
F.~Shen, Y.~Guo, C.~Du, X.~Chen, and K.~Yu, ``Acoustic bpe for speech generation with discrete tokens,'' in \emph{ICASSP 2024}, 2024, pp. 11\,746--11\,750.

\bibitem{nakamura2025discretespeechunitextraction}
T.~Nakamura, K.~Choi, K.~Hojo, Y.~Bando, S.~Fukayama, and S.~Watanabe, ``Discrete speech unit extraction via independent component analysis,'' in \emph{SALMA: Speech and Audio Language Models - Architectures, Data Sources, and Training Paradigms, IEEE International Conference on Acoustics, Speech, and Signal Processing Workshops}, 2025.

\bibitem{gat-etal-2023-augmentation}
\BIBentryALTinterwordspacing
I.~Gat, F.~Kreuk, T.~Anh~Nguyen, A.~Lee, J.~Copet, G.~Synnaeve, E.~Dupoux, and Y.~Adi, ``Augmentation invariant discrete representation for generative spoken language modeling,'' in \emph{Proceedings of the 20th International Conference on Spoken Language Translation (IWSLT 2023)}, E.~Salesky, M.~Federico, and M.~Carpuat, Eds.\hskip 1em plus 0.5em minus 0.4em\relax Toronto, Canada (in-person and online): Association for Computational Linguistics, Jul. 2023, pp. 465--477. [Online]. Available: \url{https://aclanthology.org/2023.iwslt-1.46/}
\BIBentrySTDinterwordspacing

\bibitem{vashishth2024stabspeechtokenizerassessment}
\BIBentryALTinterwordspacing
S.~Vashishth, H.~Singh, S.~Bharadwaj, S.~Ganapathy, C.~Asawaroengchai, K.~Audhkhasi, A.~Rosenberg, A.~Bapna, and B.~Ramabhadran, ``Stab: Speech tokenizer assessment benchmark,'' 2024. [Online]. Available: \url{https://arxiv.org/abs/2409.02384}
\BIBentrySTDinterwordspacing

\bibitem{yeh2024estimating}
S.-L. Yeh and H.~Tang, ``Estimating the completeness of discrete speech units,'' in \emph{2024 IEEE Spoken Language Technology Workshop (SLT)}, 2024, pp. 415--422.

\bibitem{zeghidour2021soundstream}
N.~Zeghidour, A.~Luebs, A.~Omran, J.~Skoglund, and M.~Tagliasacchi, ``Soundstream: An end-to-end neural audio codec,'' \emph{IEEE/ACM Transactions on Audio, Speech, and Language Processing}, vol.~30, pp. 495--507, 2021.

\bibitem{fossez2023high}
\BIBentryALTinterwordspacing
A.~D{\'e}fossez, J.~Copet, G.~Synnaeve, and Y.~Adi, ``High fidelity neural audio compression,'' \emph{Transactions on Machine Learning Research}, 2023, featured Certification, Reproducibility Certification. [Online]. Available: \url{https://openreview.net/forum?id=ivCd8z8zR2}
\BIBentrySTDinterwordspacing

\bibitem{shi2024espnet}
J.~Shi, J.~Tian, Y.~Wu, J.-w. Jung, J.~Q. Yip, Y.~Masuyama, W.~Chen, Y.~Wu, Y.~Tang, M.~Baali \emph{et~al.}, ``Espnet-codec: Comprehensive training and evaluation of neural codecs for audio, music, and speech,'' in \emph{2024 IEEE Spoken Language Technology Workshop (SLT)}.\hskip 1em plus 0.5em minus 0.4em\relax IEEE, 2024, pp. 562--569.

\bibitem{wu-etal-2024-codec}
\BIBentryALTinterwordspacing
H.~Wu, H.-L. Chung, Y.-C. Lin, Y.-K. Wu, X.~Chen, Y.-C. Pai, H.-H. Wang, K.-W. Chang, A.~Liu, and H.-y. Lee, ``Codec-{SUPERB}: An in-depth analysis of sound codec models,'' in \emph{Findings of the Association for Computational Linguistics: ACL 2024}, L.-W. Ku, A.~Martins, and V.~Srikumar, Eds.\hskip 1em plus 0.5em minus 0.4em\relax Bangkok, Thailand: Association for Computational Linguistics, Aug. 2024, pp. 10\,330--10\,348. [Online]. Available: \url{https://aclanthology.org/2024.findings-acl.616/}
\BIBentrySTDinterwordspacing

\bibitem{guo2024vec2wav}
Y.~Guo, Z.~Li, J.~Li, C.~Du, H.~Wang, S.~Wang, X.~Chen, and K.~Yu, ``vec2wav 2.0: Advancing voice conversion via discrete token vocoders,'' \emph{arXiv preprint arXiv:2409.01995}, 2024.

\bibitem{osakuade2024discrete}
O.~Osakuade and S.~King, ``Do discrete self-supervised representations of speech capture tone distinctions?'' \emph{arXiv preprint arXiv:2410.19935}, 2024.

\bibitem{kharitonov-etal-2022-text}
\BIBentryALTinterwordspacing
E.~Kharitonov, A.~Lee, A.~Polyak, Y.~Adi, J.~Copet, K.~Lakhotia, T.~A. Nguyen, M.~Riviere, A.~Mohamed, E.~Dupoux, and W.-N. Hsu, ``Text-free prosody-aware generative spoken language modeling,'' in \emph{Proceedings of the 60th Annual Meeting of the Association for Computational Linguistics (Volume 1: Long Papers)}, S.~Muresan, P.~Nakov, and A.~Villavicencio, Eds.\hskip 1em plus 0.5em minus 0.4em\relax Association for Computational Linguistics, May 2022, pp. 8666--8681. [Online]. Available: \url{https://aclanthology.org/2022.acl-long.593}
\BIBentrySTDinterwordspacing

\bibitem{shi2023}
A.~Sicherman and Y.~Adi, ``Analysing discrete self supervised speech representation for spoken language modeling,'' in \emph{ICASSP 2023 - 2023 IEEE International Conference on Acoustics, Speech and Signal Processing (ICASSP)}, 2023, pp. 1--5.

\bibitem{deseyssel23_interspeech}
M.~{de Seyssel}, M.~Lavechin, H.~Titeux, A.~Thomas, G.~Virlet, A.~S. Revilla, G.~Wisniewski, B.~Ludusan, and E.~Dupoux, ``Prosaudit, a prosodic benchmark for self-supervised speech models,'' in \emph{Interspeech 2023}, 2023, pp. 2963--2967.

\bibitem{seyssel-etal-2024-emphassess}
\BIBentryALTinterwordspacing
M.~de~Seyssel, A.~D{'}Avirro, A.~Williams, and E.~Dupoux, ``{E}mph{A}ssess : a prosodic benchmark on assessing emphasis transfer in speech-to-speech models,'' in \emph{Proceedings of the 2024 Conference on Empirical Methods in Natural Language Processing}, Y.~Al-Onaizan, M.~Bansal, and Y.-N. Chen, Eds.\hskip 1em plus 0.5em minus 0.4em\relax Miami, Florida, USA: Association for Computational Linguistics, Nov. 2024, pp. 495--507. [Online]. Available: \url{https://aclanthology.org/2024.emnlp-main.30/}
\BIBentrySTDinterwordspacing

\bibitem{maimon2025salmon}
G.~Maimon, A.~Roth, and Y.~Adi, ``Salmon: A suite for acoustic language model evaluation,'' in \emph{ICASSP 2025-2025 IEEE International Conference on Acoustics, Speech and Signal Processing (ICASSP)}.\hskip 1em plus 0.5em minus 0.4em\relax IEEE, 2025, pp. 1--5.

\bibitem{morise2016world}
M.~Morise, F.~Yokomori, and K.~Ozawa, ``World: a vocoder-based high-quality speech synthesis system for real-time applications,'' \emph{IEICE TRANSACTIONS on Information and Systems}, vol.~99, no.~7, pp. 1877--1884, 2016.

\bibitem{nguyen-etal-2023-generative}
\BIBentryALTinterwordspacing
T.~A. Nguyen, E.~Kharitonov, J.~Copet, Y.~Adi, W.-N. Hsu, A.~Elkahky, P.~Tomasello, R.~Algayres, B.~Sagot, A.~Mohamed, and E.~Dupoux, ``Generative spoken dialogue language modeling,'' \emph{Transactions of the Association for Computational Linguistics}, vol.~11, pp. 250--266, 2023. [Online]. Available: \url{https://aclanthology.org/2023.tacl-1.15/}
\BIBentrySTDinterwordspacing

\bibitem{qian2022contentvec}
K.~Qian, Y.~Zhang, H.~Gao, J.~Ni, C.-I. Lai, D.~Cox, M.~Hasegawa-Johnson, and S.~Chang, ``Contentvec: An improved self-supervised speech representation by disentangling speakers,'' in \emph{International conference on machine learning}.\hskip 1em plus 0.5em minus 0.4em\relax PMLR, 2022, pp. 18\,003--18\,017.

\bibitem{baevski2022data2vec}
A.~Baevski, W.-N. Hsu, Q.~Xu, A.~Babu, J.~Gu, and M.~Auli, ``Data2vec: A general framework for self-supervised learning in speech, vision and language,'' in \emph{International conference on machine learning}.\hskip 1em plus 0.5em minus 0.4em\relax PMLR, 2022, pp. 1298--1312.

\bibitem{ma2024icassp}
Z.~Ma, W.~Wu, Z.~Zheng, Y.~Guo, Q.~Chen, S.~Zhang, and X.~Chen, ``Leveraging speech ptm, text llm, and emotional tts for speech emotion recognition,'' in \emph{ICASSP 2024 - 2024 IEEE International Conference on Acoustics, Speech and Signal Processing (ICASSP)}, 2024, pp. 11\,146--11\,150.

\bibitem{ma-etal-2024-emotion2vec}
\BIBentryALTinterwordspacing
Z.~Ma, Z.~Zheng, J.~Ye, J.~Li, Z.~Gao, S.~Zhang, and X.~Chen, ``emotion2vec: Self-supervised pre-training for speech emotion representation,'' in \emph{Findings of the Association for Computational Linguistics: ACL 2024}, L.-W. Ku, A.~Martins, and V.~Srikumar, Eds.\hskip 1em plus 0.5em minus 0.4em\relax Bangkok, Thailand: Association for Computational Linguistics, Aug. 2024, pp. 15\,747--15\,760. [Online]. Available: \url{https://aclanthology.org/2024.findings-acl.931/}
\BIBentrySTDinterwordspacing

\bibitem{libri}
V.~Panayotov, G.~Chen, D.~Povey, and S.~Khudanpur, ``Librispeech: An {ASR} corpus based on public domain audio books,'' in \emph{ICASSP 2015}, 2015, pp. 5206--5210.

\bibitem{onda2025discretetokensexhibitinterlanguage}
K.~Onda, K.~Imoto, S.~Fukayama, D.~Saito, and N.~Minematsu, ``{Discrete Tokens Exhibit Interlanguage Speech Intelligibility Benefit: an Analytical Study Towards Accent-robust ASR Only with Native Speech Data},'' in \emph{{Interspeech 2025}}, {2025}, pp. {221--225}.

\bibitem{mead}
\BIBentryALTinterwordspacing
K.~Wang, Q.~Wu, L.~Song, Z.~Yang, W.~Wu, C.~Qian, R.~He, Y.~Qiao, and C.~C. Loy, ``Mead: A large-scale audio-visual dataset for emotional talking-face generation,'' in \emph{Computer Vision ^^e2^^80^^93 ECCV 2020: 16th European Conference, Glasgow, UK, August 23^^e2^^80^^9328, 2020, Proceedings, Part XXI}.\hskip 1em plus 0.5em minus 0.4em\relax Berlin, Heidelberg: Springer-Verlag, 2020, p. 700^^e2^^80^^93717. [Online]. Available: \url{https://doi.org/10.1007/978-3-030-58589-1_42}
\BIBentrySTDinterwordspacing

\bibitem{kando2025exploringeffectsegmentationvocabulary}
S.~Kando, Y.~Miyao, and S.~Takamichi, ``{Exploring the Effect of Segmentation and Vocabulary Size on Speech Tokenization for Speech Language Models},'' in \emph{{Interspeech 2025}}, {2025}, pp. {5728--5732}.

\bibitem{POEPPEL2003245}
\BIBentryALTinterwordspacing
D.~Poeppel, ``The analysis of speech in different temporal integration windows: cerebral lateralization as ‘asymmetric sampling in time’,'' \emph{Speech Communication}, vol.~41, no.~1, pp. 245--255, 2003, the Nature of Speech Perception. [Online]. Available: \url{https://www.sciencedirect.com/science/article/pii/S0167639302001073}
\BIBentrySTDinterwordspacing

\bibitem{timit}
J.~Garofolo, L.~Lamel, W.~Fisher, J.~Fiscus, D.~Pallett, N.~Dahlgren, and V.~Zue, ``Timit acoustic-phonetic continuous speech corpus,'' \emph{Linguistic Data Consortium}, 11 1992.

\bibitem{onda2025differentiablekmeansfullyoptimizeddiscrete}
K.~Onda, Y.~Kashiwagi, E.~Tsunoo, H.~Futami, and S.~Watanabe, ``{Differentiable K-means for Fully-optimized Discrete Token-based ASR},'' in \emph{{Interspeech 2025}}, {2025}, pp. {1223--1227}.

\bibitem{yoon23c_interspeech}
J.~W. Yoon, S.~M. Kim, and N.~S. Kim, ``Mcr-data2vec 2.0: Improving self-supervised speech pre-training via model-level consistency regularization,'' in \emph{Interspeech 2023}, 2023, pp. 2833--2837.

\end{thebibliography}

\end{document}